\documentstyle[12pt]{article} 
\begin{document} \baselineskip 0.7cm

\begin{flushright} ULB-TH-97/05  \end{flushright} \begin{center}

\vskip2cm {\large \bf Notes on Connes' Construction of the Standard Model}

(Based on a talk presented at the San Miniato meeting
 
``The Inevitable Rise of the Standard Model", April 21-25, 1997)

\vskip2cm R. BROUT\footnote{e-mail: rlareppe@ulb.ac.be} \vskip2cm Service de
Physique Th\'eorique\\

Universit\'e Libre de Bruxelles, Campus de la Plaine, C.P. 225 \\

Boulevard du Triomphe, B-1050 Bruxelles, Belgium \end{center}

\vskip3cm \noindent {\bf Abstract} \bigskip

The mathematical apparatus of non commutative geometry and operator algebras 
which Connes has brought to bear to construct a rational scheme for the
internal symmetries of the standard model is presented from the physicist's
point of view. Gauge symmetry, anomaly freedom, conservation of electric
charge, parity violation and charge conjugation all play a vital role. When
put together with a relatively simple set of algebraic algorithms they
deliver many of the features of the standard model which otherwise seem
rather ad hoc.

\newpage

\section{Introduction} \bigskip

In recent years the approach of Connes$^{1), 2), 3), 4), 5), 6)}$, using
techniques of operator algebra and non-commutative geometry, has given new
insights into the symmetry structure of the standard model (SM). Many of the
otherwise inexplicable features of SM have been fitted into a neat rational
scheme,  the success of which suggests that Connes' methods may well
constitute one of the ingredients of the physics of the future. This is
particularly true since the geometric concepts that are used are
generalizations of usual continuum geometry to discrete non-commutative
spaces. And this seems to be where quantum gravity is pointing to. \bigskip

As Connes' works themselves are couched in mathematical terms which generally
are not part of the baggage of the everyman physicist, several expos\'es have
been written whose purpose is to bridge the information gap. The present paper
is conceived in the same vein, but with still more emphasis on the physics,
and with less use of notation and language that is unfamiliar to physicists.
My aim is to draw the attention of a larger part of the physics community to
this development than has been the case heretofore.

\bigskip

My plan is take up severally, certain peculiar features of SM, mostly
concerned with internal symmetries, and to show how Connes' methods deal with
them. I emphasize strongly the gauge principle and its handmaiden, anomaly
freedom. One of the more remarkable things that has come up as the uncanny
consistency of Connes' approach with the constraints of anomaly freedom.

\bigskip

From the above it is clear that what follows will be very far from a
systematic presentation. For that the reader may consult the very fine
expos\'es on the subject. Furthermore their existence makes it pointless to
include calculational detail in the present survey. Rather I sketch the ideas
behind each point I take up and then refer to this or that reference which I
esteem detailed and pedagogical concerning the item under consideration.
Obviously nothing in the foregoing is original, if not perhaps a fresh
appreciation of certain points.

\newpage \section{Why Fundamental Representation}

\bigskip

In SM, leptons and quarks serve as representations of non-Abelian groups.
These reps are the fundamentals. Why? \bigskip

The basic tool of Connes is the action of algebraic elements, built on points
of a manifold, on a Hilbert space, ${\cal H}$, the latter being spanned by the
elementary fermions. Internal symmetries are developed through use of the
algebras $M_n (C) = n \times n$ matrices whose entries are complex numbers.
(For $ n = 2$, there is an important subalgebra of $M_2$, the quatermons,
$\mid\!$H, of the form $ A + i \stackrel{\rightharpoonup}{B}.
\stackrel{\rightharpoonup}{ \sigma}$ with $ A, \stackrel{\rightharpoonup}{ B}$
real and $ \stackrel{\rightharpoonup}{ \sigma}$ the Pauli matrices and for $
n = 1$, the algebra is that of the complex numbers ${\bf C}$.). Unlike groups,
algebras have the additional property of linearity: if $a, b$ are elements of
the algebra then $ c \in {\cal A}$ if $ c = \lambda_1 a + \lambda_2 b$ with $
\lambda_1, \lambda_2 \in {\bf C}$ and $ a, b \in {\cal A}$ in addition to the
group property $ c = a b$. This linearity restricts representations to the
fundamental -one checks that tensor products reps of $M_n (C)$ violate
linearity. \bigskip

Lest there be any misunderstanding the linearity property does admit tensor
products of different algebras and indeed in Section 6 we shall consider
algebras which are sums of simple algebras, flavor + color. Members of ${\cal
H}$ can have both attributes i.e. be tensor products. But the action of each
subalgebra on the appropriate index of a component of such a representation is
necessarily in the fundamental in order that \underline{it} be an algebra.
This property, in itself, to my mind is quite sufficient to motivate the
algebraic approach.

\vskip1cm \section{Fermionic Mass, Spontaneously Broken Chiral Symmetry and
Parity Violation  }

\bigskip

It is universally granted that one of the marvels of SM is the simultaneous
occurrence of gauge symmetry governing the dynamics of all interacting vector
and axial currents, and the existence of massy fermions. The reconciliation
is, of course, effected by the mechanism of spontaneous broken chiral
symmetry (S B $\chi$ S). And, of course, it is the gauge symmetry that makes
the theory renormalizable, as was anticipated shortly after the discovery of
massy gauge mesons$^{ 7)}$ and subsequently proven in the important works of
Veltman and 't Hooft $^{8)}$.

\bigskip

SB$\chi$S, in the context of gauge theory, was developed simultaneously and
independently by Englert and myself $^{9)}$ and by Higgs $^{10)}$. In the
former work two possibilities were considered: 1) the scalar responsible for
SB$\chi$S was expressed in terms of a dynamically generated composite field,
as proposed in prior work of Nambu or 2) it was an elementary scalar field;
Higgs only considered this latter possibility. The issue is still not
settled. In Connes'approach, SB$\chi$S is realized in terms of an elementary
scalar field. But it is wise to bear in mind that the theory so obtained, it
is wise to bear in mind that the theory so obtained is ``effective" i.e.
thought to be valid at a certain length scale. For an example of how this
might be constructed see ref.$^{6)}$. It is not excluded at a smaller scale
that new phenomena will occur which would lead to compositeness, such as
encountered in efforts to make SB$\chi$S a dynamical theory. Thus I prefer to
reserve judgement and consider Connes'approach, at its present stage, to be
phenomenological.

\bigskip

Now for one of Connes' main ideas. Scalars are gauge bosons that serve as
connections in a manifold comprised of 2 points: left (L) and right (R), which
I now explain at length. \bigskip

Fermions, in the absence of mass, are displaced in space-time, on two
different surfaces L and R, through the action of the Dirac operator. The
usual Yang-Mills fields (YM) supply connections on these surfaces so as to
allow internal symmetries to be gauged. The vector and axial currents coupled
to the YM are sums of bilinears in L \underline{or} R fermions and do not mix
L \underline{and} R. i.e. the coupling keeps an L (R) fermion on the L (R)
surface. The role of mixing L and R is taken on by the scalars $\phi$ through
a coupling ($\psi_L ^+ \psi_R \phi + h.c.$). The bilinear $ \psi_L^+ \psi_R$ 
is a sort of current between the 2 corresponding points L and R at the same
space-time point $x^\mu, \phi$ being the ``gauge scalar". The thought is not
only engaging, but it ties in most elegantly with Connes' rewriting of
geometry on discrete spaces in terms of axioms which permit a natural
generalization of all continuum concepts. These discrete spaces can be
composed of but 2 points -L and R in particular, and so are applicable to SM
gauge geometry. \bigskip

To implement this extended sense of the gauge principle Connes defines
generalized differentiation from the commutator with the Dirac operator, D. In
L, R representation one has for a single fermion of mass, M

\begin{eqnarray} \label{1} D = D_0 + D_M = \left( \begin{array}{cc} i
\partial\!\!\!/ & 0 \\ 0 & i \partial\!\!\!/  \end{array}\right) +
\left(\begin{array}{cc} 0 & M \\ M & 0  \end{array}\right) \end{eqnarray}
\bigskip

\noindent where each entry is a diagonal 2 $ \times$ 2 matrix. Usual
differentiation of a function $ f (x)$, now regarded as the representative of
an algebraic element on the point, $x$, of the space-time manifold, is
obtained from $ [ D_0, f] = i \partial\!\!\!/ f$. In this way one can proceed
systematically to invent the whole system of exterior derivatives so as to
recover Riemannian geometry on one hand, and usual Yang-Mills theory on the
other, when $f$ is replaced by matrices of functions. This is \underline{not}
the subject of this paper. Rather we have introduced this notion of
differentiation to motivate the construction of the ``derivative" in the space
of 2 points L, R through

\begin{eqnarray} \label{2} \delta a = [ D_M, a ]\; . \end{eqnarray}

\bigskip

\noindent $ D_M$ being L, R non diagonal, thereby gives the sense of
displacement L $\leftrightarrow$ R. If there is more than one fermion in
${\cal H}$ than $a$ is a representative of some matrix algebra, to be
specified, that is erected on the points L and R. And M is then a matrix
whose structure reflects internal symmetries. For example for quarks $u, d$
of one generation M is a diagonal matrix with elements $M_u, M_d$. In the
presence of more than one generation, mixing is expressed through
Kobabayashi-Maskawa mixing in the matrix $M_d$. What is of primary importance
is charge conservation.  There is no $ u, d$ mixing in the matrix $M$.
Axiomatically, there is a charge operator $Q$ such that $ [ D, Q] = 0$. This
will be used extensively in what follows.

\bigskip

As has been stated, the algebra representative $\rho (a)$ is erected on the
points L and R. $\rho (a)$ comes in two blocks $ \rho_L (a)$ and $\rho_R(a)$,
corresponding to the 2 points of the manifold. Axiomatically, there is an
operator $\chi$ such that $ \chi^2 = 1$. The action of $ \chi$ on ${\cal H}$
gives + 1 on components in the L sector and - 1 on the R sector. $ \chi$
anticommutes with $D_M$ and commutes with $a$ (for $ a \in {\cal A}$). This is
enough to give the representations

\begin{eqnarray} \label{3} \rho(a) = \left( \begin{array}{cc} \rho_L (a) & 0
\\ 0 & \rho_R (a) \end{array} \right) ;\;  D_M = \left( \begin{array}{cc} 0 &
M \\ M^+ & 0 \end{array} \right) ; \; \chi = \left( \begin{array}{cc} I & 0 \\
0 & - I \end{array} \right)  \end{eqnarray}

\bigskip

\noindent where, as previously stipulated $ [ D_M, Q] = 0$, so $M$ has no non
diagonal elements between different charge states. \bigskip

This last point warrants a digression into the underlying physics. In SM it is
supposed that one symmetry is left unbroken. This is a natural condition in
that the induced mass term is of the form $ \psi_R \psi_L^+ \langle \Phi
\rangle$ (where $ \langle \Phi \rangle$ is the vacuum expectation value of a
scalar or pseudoscalar field). This has as consequence a symmetry wherein,
for a given component, $\psi_R$, and its counterpart, $\psi_L$, one
multiplies each by the same phase factor - a universal phase angle multiplied
by the electric charge of the component in question. This phase is the
electromagnetic phase and the unbroken symmetry results in one gauge field
retaining zero mass. The Hilbert space can thus be classed by eigenfunctions
of this charge, $Q$, and this must be respected by the dynamics. Connes has
incorporated this physical requirement into the formalism through $ [ D_M, Q]
= 0$.

\bigskip

An interesting aside that excites the imagination: in the geometric
formulation of this concept of differentiation, Connes has generalized the
notion of distance between 2 points of a manifold. For a single fermion at
fixed $x^\mu$, the distance between L and R is $ M^{-1}$.

\bigskip

Now to internal symmetries and the gauge principle. Because a Hilbert space
exists one can invent an action $ \psi^+ D \psi$ and one is lead to
investigate its invariances under automorphisms of the algebra. The
components of $\psi$ are in ${\cal H}$. Henceforth unless otherwise specified
we deal only with the internal symmetries so that $D$ will be the internal
part, $ D_M$, only.

\bigskip

The bilinear $ \psi^+ \psi$ is invariant under the subensemble of ${\cal A}$
which is its set of unitary matrices. For our example this set is represented
by the group $ U_L \times U_R$ i.e.

\begin{eqnarray} \label{4} \rho (u) = \left( \begin{array}{cc} u_L & 0 \\ 0 &
u_R \end{array} \right) \end{eqnarray} \bigskip

\noindent with $ u_L \in U_L, u_R \in U_R$ (For quaternions $ U_L$ is not
$U_2$, but $ S U_2$ and from now on when the internal space is 2 dimensional
we will work with $\mid\!$H  and not $ M_2 (C)$. In that case, the group is
i-spin. This choice is made because it tallies correctly with the observed
hypercharge assigments. (See Sections 5 and 6.)

\bigskip

If $ \rho_L \neq \rho_R$, then $ [ D_M, \rho] \neq 0$, whence $ \psi^+ D_M
\psi$ is not invariant under transformation of $\psi$ by the unitaries.
Following time honored procedure, we therefore introduce ``vector or gauge
potentials" to construct an invariant action beginning with $\psi^+ D_M \psi$.
Since $ \psi \to U \psi$ induces the change

\begin{eqnarray} \label{5} \delta \psi^+ D_M \psi = \psi^+ U^+ [ D_M, U] \psi
\; , \end{eqnarray}

\bigskip

\noindent Connes invents a gauge potential of the form $ \sum \alpha_i [ D,
\alpha_i ^\prime ]$ with $ \alpha_i, \alpha_i^\prime \in {\cal A}$ and for
internal symmetries these elements are decomposed into $ \rho_L$ and $ \rho_R$
of Eq. 3. Since $ U^+ [D_M, U]$ is a hermetian form the gauge potentials are
taken to be hermetian as well their role being to compensate for the variation
5).  Indeed  under $ \psi \to U \psi$ one requires the gauge potential $V$ to
transform as

\begin{eqnarray} \label{6} V \to UVU^+ - U^+ [D, U] \end{eqnarray}

\bigskip

\noindent thereby ensuring the invariance of the action $ A$ under unitary
transformation, where

\begin{eqnarray} \label{7} A = \psi^+ [ D + V ] \psi \end{eqnarray}

\bigskip \noindent $V$'s which transform like 6) are of the form $ \sum_i a_i
[D, b_i]$ which are hermetian; $a_i, b_i \in {\cal A}$.

\bigskip

\noindent [I have explicitly not used $D_{M}$ but $D$ in Eq. (7) in order to
suggest to the reader to carry out the steps to recover $Y M$ potentials in
the case where $ D = D_0$ and $U$ is taken to depend on the space-time
variables $x^\mu$. He may also wish to construct the covariant fields $F_{\mu
\nu}$. For this consult refs $^{1) 2), 11), 12)}$.

\bigskip

The law of transformation (6) has two parts. The first term, covariant, is
simply a unitary transformation of all the members of the algebra -a one to
one mapping to which corresponds the unitary transformation of ${\cal H}$. The
second term is not covariant. Connes says that $ D + V$ corresponds to a
deformation of the space -a modification of its metric properties. This is
understood from the fact that the operator $D$ (which is after all a sort of
square root of a laplacien) encodes this metric. One may think of the
deformation of $\partial\!\!\!/$ by gravity in the vierbein formulation.

\bigskip

For the case $ D = D_M$, the gauge potentials are thus sums of terms of the
form

\begin{eqnarray} \label{8} V = \left( \begin{array}{cc} \rho_L^\prime & \; \\
 \;  & \rho_R^\prime \end{array} \right) \left( \begin{array}{cc} 0 & \rho_L
M- M \rho_R \\ M^+ \rho_L - \rho_R M^+ & 0 \end{array} \right) + h.c.
\end{eqnarray}

\bigskip

As an example drawn from the Glashow Weinberg Salam scheme  (G W S) for
quarks one picks $(\rho_L \in \vert\!{\rm H}, \rho_R $ = diag $ (\lambda,
\lambda^*)$ with $ \lambda \in C$. Correspondingly $ u_L \in S U_2, u_R$ =
diag $(e^{i \varphi}, e^{-i \varphi})$. The Hilbert space is $ u_L, d_L$ (an
i-spin doublet) and $u_R, d_R$ (singlets). A typical term in $V$ is
proportional to $ q M$ where  $ q \in \vert\!{\rm H}$. That this is an
appropriate representation of the scalar potential of the G W S scheme is
well known. For details see ref. $^{12)}$.

\bigskip

The point I want to stress here is that non trivial dynamics arises only in
virtue of $[ D_M, \rho]Ê\neq 0$. This excludes a pure vector theory with
unbroken symmetry, like QED or QCD wherein $\rho_L = \rho_R$ and $M$
proportional to $I$. The physically realized theory, SB$\chi$S with L and R
represented by distinct algebras fits nicely into Connes' algorithm. But this
method does not exclude parity non-violating vector theories with $M$  having
different diagonal enties. An example is where the unitaries are $(S U_2)_L
\times (S U_2)_R)$. So Connes algorithm only goes part of the way towards
picking out the correct theory. More detail on this point is in ref.$^{13)}$.

\bigskip

It is important to be conscious of the novel feature of Connes'algorithm
wherein the dynamics is postulated in terms of the c-number entries of the
matrix $M$. In usual gauge theory one introduces a scalar field, an operator
with certain quantum numbers (or possibly a non-local construction as
envisaged in theories of dynamically broken symmetry.) There is no question of
postulating commutation properties for the expectation value of that field.
They emerge, in principle from the field theory. Connes reverses the order. He
handles masses as inputs -and to this extent what he does is low energy
phenomenology, building on the form of an effective theory as it were. Thus
where the field theorist would have no objection, \`a priori, to SB$\chi$S in
QCD or QED as conventionally formulated, Connes effectively discards this
possibility at the outset. The response of some physicists is then to discard
the theory as well. And others will play along once noticing that
Connes'algorithms, come out systematically on top. This seems to be the way
nature works. Such physicists will then start to ask why. It is purpose here
to bring this option to his attention.

\bigskip

To see how SB$\chi$S in Connes'approach comes about one derives an action for
$q$ by following the YM analogy. ``Exterior derivatives" which are the
analogies of $\partial_\mu A_\nu - \delta_\nu A_\mu$ are found from the rule $
\delta (a_i [D, b_i] ) = \delta (a_i \delta b_i) = [D, a_i] [D, b_i] = [\delta
a_i \delta b_i]$. This rule is dictated by the construction of the algebra of
exterior forms wherein $Ê\delta^2 = 0$. (For those unfamiliar with this
concept an introduction is to be found in ref. $^{14)}$ and more formally in
ref. $^{15)}$. The covariant field, often called the curvature $C$, is
constructed from $ \delta V + V \times V$. [It is a nice exercice to prove
that this form is covariant in consequence of (6)].

\bigskip

$C$ being covariant, $tr C^2$ is invariant and from the hermetian character of
$V$ it is easy to prove that this form is non negative. The trace is over the
Hilbert space indices. For our GWS example one finds $tr C^2$ proportional to
$ tr [ \vert q + 1 \vert^2 - 1 ]^2$, therefore having a minimum at $q = 0$. It
is assumed that the action is $ tr C^2$ or a sum of positive powers thereof.
This latter possibility is not generally considered in analogy to the Y M
action being $tr F^2$. But whereas this latter is supported by a
renormalization group (RG) analysis wherein ``irrelevant terms" are shown,
indeed, to be irrelevant, there is no such calculation carried out in the
present case. The proposition of ref. $^{6)}$ requires further investigation
in this regard. It may lead to interesting insights into usual RG analyses and
quantum gravity.

\bigskip

The reader may be perplexed that SB$\chi$S is encoded in the minimum at $q =
0$. But this is as it should be. The fermion action is $\psi^+ [ M + V]
\psi$, which in our case takes on forms like $\psi_R ^+ M (q + 1) \psi_L$.
Therefore the scalar field $\Phi$ is proportional to $q+1$, since the Yukawa
couplings are contained in the components of $M$. Precisely one has $ q + 1 =
(\Phi / < \Phi > )$. The phenomenom of S B$\chi$S is $M \neq 0$. Fluctuations
around $M$ are thus encoded in $q$ and $<q> = 0$ in vacuum. Moreover, from 7,
one has that $D_M + V$ is covariant, (by construction), so that $\Phi$ is
covariant. Since $q+1$ and $\Phi$ are related by multiplication by $M$, it is
then no surprise that the effective potential $(= tr C^2)$ is expressible in
powers of $(q+1) \times$ powers of masses. More detail on the relation of $q$
to $\Phi$ and $M$ is to be found in refs $^{11), 12)}$. \bigskip

Generalization to leptons is obtained by eliminating $u_R$ and setting $m_u =
0$. Many generations are handled by extending ${\cal H}$ including KM mixing
the $d^\prime s$.

\bigskip

The above internal algebraic constructions are then combined with space-time
through use of the part of $D$ equal to $i \partial \!\!\!/$. Components of
${\cal H}$ now depend on $x^\mu$ as well. One follows the YM procedure in its
Connesian algebraic guise. At the (quadratic + quartic) level one finds for
the bosonic action what one expects, a sum of the gauge invariant terms, the
$Y M$ action $ (= tr F^2)$, the kinetic term of the scalar properly
covariantized $ =(D_\mu \Phi)^2$ (where $ \Phi = q + 1$ for the G W S
example) and a term proportional to $tr C^2$. An interesting complication
arises with respect to this latter, to wit:

\vskip1cm

\section{Why More Than One Generation} \bigskip

All of these algebraic manipulations involve homomorphisms of the algebras
(internal and space-time) on to their representations (matrices and functions
of $ x^\mu$). One encounters kernels; more precisely $\delta$(Kernel) is non
zero. These must be divided out. As an example, a term arises in the second
order exterior derivative which is represented at the first order level by
zero. Thus, physically, one would get a field from a zero potential. This
term is non gauge invariant and must be eliminated.

\bigskip

The interesting point that arises is that, upon tensoring the internal and
external algebras, the two kernels intersect. The external algebra gives rise
to a kernel which has a piece of the second order exterior derivative which is
the algebra itself. In consequence, upon quotientizing one loses the term in
$tr C^2$ when there is only one generation. In short, following the algebraic
rules results in the loss of SB$\chi$S when there is only one generation.
Then, eureka, it comes back for more than one generation and once more yields
a non-negative effective potential in $\vert \Phi \vert^2$ of minimum $ \Phi
\neq 0$. For ample detail on this calculation a good reference is ref.
$^{11)}$.

\bigskip

As far as I know this is the only place in standard model physics where a
rationale has been supplied to this otherwise inexplicable physical phenomenom
of the existence of several copies of the same representation. Conne's
algorithms ties non trivial dynamics to their existence, but at present do
not explain why there are just three.

\bigskip

Let us now return to the main line of interest of this survey -the
characterization of internal symmetries in terms of Connes' constructions.

\bigskip

In the above GWS example there has been no question of hypercharge. The $ u_L,
d_L$ (or $ e_L, \nu_L)$ doublet thus gives an electric charge splitting of the
doublet into equal and opposite values. The principle of charge conservation 
$([D, Q] = 0)$ then forces $\rho_R$ to be of the form diag
$(\lambda,\lambda^*)$ as well) as we have postulated above -since the R
fermions must be split in electric charge just as their L partners. It is the
unitary part of $ C (= U_1)$ which in the R algebra is represented by $U_R
\in U_1 \times U_1 \ (u_R : {\rm diag} (e^{i \varphi}, e^{- i \varphi})$
which does the job.

\bigskip

This version of GWS is at odds with observation and the scheme must be
modified. I shall discuss this elaboration, first from the point of view of
anomaly freedom (Section 5) and then present Connes's construction of
bimodules whereupon it will be seen how nicely they fit together to lead to
the structure of the observed SM.

\vskip1cm

\section{Anomaly Freedom} \bigskip

Anomalies arise from the sum of the  vertex corrections, of the axial
currents $^{16)}$. If they do not vanish the current coupled to the $
\gamma_\mu \gamma_5$ vertex is not conserved; gauge invariance is violated and
the theory is not renormalizable. In brief, consistency requires vanishing of
the anomaly. [Note that the above remarks apply to those axial currents which
are coupled to gauge fields. Sometimes in physics one encounters anomalous
currents which are not so coupled, such as in the famous $U(1)$ problem of P
C A C.]

\bigskip

We shall begin with the lepton sector, assuming some of the elements of the G
W S scheme. Since the latter is anomalous, fermions other than leptons must
exist. We introduce quarks and show how they manage to cancel the anomaly. I
shall not discuss hypothetical schemes based on GUTS. They seem to have no
basis in observation. What is remarkable is that there is one simple scheme,
the observed one, that does work. And that this scheme follows in great
measure from Connes' axioms concerning real algebraic structures.

\bigskip

Several important results follow from the anomaly free G W S scheme (an $e_L,
\nu_L$ i-spin doublet and $e_R$ singlet) taken together with the simplest and
most natural i-spins of quarks $(u_L d_L$ doublet and $u_R, d_R$ singlets).
Firstly one finds that the neutrino has electric charge zero $(Q_\nu = 0)$.
The number of colors, $C$, is not determined, but one does find that if $C$ is
odd, a collection of $C$ quarks-which is what is required to make a hadron, a
totally antisymmetric color state, has integer charge. It is then always
possible to build a ``proton", a hadron of equal and opposite charge to the
electron. Finally the sum of all fermion charges is zero. This is called the
unimodular condition and it played an important role in the formulation of
GUTS. This latter, as mentioned, seems not in accord with nature; the proton
refuses to decay. Nevertheless unimodularity has an interesting theoretical
status which will be discussed in Section 6. For completeness we now present a
brief deviation of these results.

\bigskip

It suffices to work with one generation since the quantum numbers of the
members of each generation are in one to one correspondence. The anomaly is
proportional to $^{16)}$.

\begin{eqnarray} \label{9} tr ^\prime \lambda_a \{ \lambda_b, \lambda_c \}_+
\end{eqnarray}

\bigskip

\noindent and this is required to vanish. Here $ t r^\prime = \sum_L -
\sum_R$, the minus sign because $ \gamma_5 = \pm 1$ for L, R resp.. The $
\lambda^\prime$s are group generators coupled to the gauge fields at each
vertex where $a$ is an axial and $b, c$ are vectors.

\bigskip

Consider leptons. Were there only $\nu_L, e_L$, an i-spin doublet, there
would be no anomaly. Nor would $e$ have a mass. For this one needs $ e_R \neq
0$. Of course one could also have $ \nu_R \neq 0$ and $ (e_R, \nu_R)$ another
doublet. Nature does not work that way. Let us then follow nature, exclude
$\nu_R$, and study the consequences. The result is that the leptonic sector is
anomalous, which is clear from (9) since L, R cannot balance out. We now
introduce hypercharge, defined through the Gell-Mann Nishijima relation $ Q =
T_3 + Y/2$, and for reasons discussed in Section 4), impose the condition
that, when a fermion exists in both L and R versions, one has $Q_L = Q_R$.
\bigskip

From the properties of i-spin, one readily sees that all combinations of $a,
b, c$ of Eq. 9) reduce to two conditions: $a = Y, b, c $= i-spin (with $\{
T_b, T_c\} = {1 \over 2} \delta_{b c}$, and $a, b, c$ all = $Y$. In the first
instance only L contributes and we find a leptonic anomaly given by

\begin{eqnarray} \label{10} \sum Y_L^{lept} = \sum Q_L^{lept} = Q_\nu + Q_e =
- 1 + 2 Q_\nu \end{eqnarray}

\bigskip

\noindent The cubic anomaly is

\begin{eqnarray} \label{11} \sum (Y_L^{lept})^3 - \sum (Y_R^{lept})^3 = 6 -
12 Q_\nu + 8 Q_\nu^3 \end{eqnarray}

\bigskip

\noindent [This follows from  $ \sum (Q_L^{lept})^3 - (Q_R ^{lept})^3 =
Q_\nu^3$ and use of the G M N relation in conjunction with (10)].

\bigskip

Since there is no $Q_\nu$ that sets (10) and (11) simultaneously to zero,
there must exist other fermions to cancel the leptonic anomalies.
\underline{The G W S leptonic scheme of itself is not} \underline{quantum
mechanically viable}. Let us assume, in accord with Nature, that the
accommodation to the leptonic anomaly is as simple as possible, to wit: there
is only one other kind of fermion, the quark. To distinguish it from leptons
it must bear another quantum number, color. Here we postulate that there are
$C$ colors which form the basis for a gauged unitary group (Q C D). 

\bigskip

The assumption that the (colored) quarks are the only additional type of
fermions implies that the color gauge group is pure vector. For otherwise
there would be an anomaly involving gluons. To cancel this would then require
other kinds of colored fermions.

\bigskip

To cancel Eq. 10, the L quarks must be an i-spin doublet $(u_L, d_L$ - each in
C versions). They must also come in an R version to avoid a color anomaly, as
well as to accommodate (11) and we postulate that there is both $u_R$ and
$d_R$. They cannot be an i-spin doublet since this would undo the
cancellation mission for which $(u_L, d_L)$ were invented. Therefore we
postulate each to be i-spin singlets. The rest is simple arithmetic.

\bigskip

\begin{eqnarray} \label{12} 0 = \sum Y_L = \sum Q_L \end{eqnarray} \bigskip

\noindent and (11) reads

\begin{eqnarray} \label{13} 0 &=& {1 \over 8} \sum (Y_L^3 - Y_R^3) = \sum
(Q_L - T_3)^3 - \sum Q_R ^3 \nonumber \\ &=& \sum (Q_L^3 - Q_R ^3) + {3 \over
4} \sum Q_L = Q_\nu ^3 \end{eqnarray}

\bigskip

\noindent The first conclusion is that the neutrino must be electrically
neutral. Quantum mechanics assures that if parity is broken, it does so most
elegantly. Thus one recovers the G W S assignments $ Y_L ^{lept} = - 1, \;
Y_R^{lept} = - 2$ for the leptons. It then follows from (12) that

\begin{eqnarray} \label{14} Y_L^{quark} &=& (1 / C) \nonumber \\ Q^{up} &=&
{1 \over 2} (1 + 1 / C) \\ Q^{down} &=& {1 \over 2} ( -1 + 1 / C) \nonumber
\end{eqnarray}

\bigskip

\noindent

\begin{eqnarray} \label{15} Y_R^{up} &=& (1 + 1/ C) \nonumber \\ Y_R^{down}
&=& (- 1 + 1 / C) \end{eqnarray} \bigskip

\noindent and one has established

\begin{eqnarray} \label{16} \sum Q_L = \sum Q_R = \sum Y_L = \sum Y_R = 0
\end{eqnarray}

\bigskip

\noindent Eq. (16) implies the unimodularity condition (i.e. $\sum (Y_L +
Y_R) = 0)$ discussed further in Section 6. In addition from (14) one sees
that a collection of $C$ quarks has integer (half integer) charge according
to $C$ odd (even). For $C$ odd, it is then always possible to form a
``proton",  color antisymmetric state composed $(C \pm 1)/ 2$ up and down
quarks respectively. Hence one can make neutral atoms.

\bigskip

A further remarkable element of consistency occurs when one includes gravity
in the game. A potential anomaly occurs when the indices $b$ and $c$ of Fig.1
are gravitons and $a$ is the hypercharge. Cancellation requires

\begin{eqnarray} \label{17} \sum Y_L - \sum Y_R = 0 \end{eqnarray} \bigskip

\noindent resulting once again in $Q_\nu = 0$.

\bigskip

Strange that the neutrality of the universe should depend on the vicissitudes
of an accommodation to a quantum anomaly.

\vskip1cm

\section{Charge Conjugation and the Construction of Bimodules}

\vskip1cm

For Connes'construction of bimodules $^{4), 12)}$ it is assumed that the
algebra has both the G W S structure of quarternions and complex numbers
$(\mid \!\!\!{\rm H} \oplus {\bf C})$, and takes into account color as well.
So for $C$ colors the algebra is $ (\mid \!\!\!{\rm H}+ {\bf C}+ M_c ({\bf
C})$. Once more in what follows it suffices to work with one generation and
the Hilbert space ${\cal H}$ is modelled after G W S: $ (e_L, \nu_L) (u_L,
d_L), e_R, u_R, d_R$ having $ 3 + 4 C$ components. The bimodular construction
is based on the fact that the complete Hilbert space is $ {\cal H} +
\overline{{\cal H}}$ where $\overline{{\cal H}}$ contains $3 + 4 C $
antiparticles. This is obtained from ${\cal H}$ by taking its C P conjugate,
the latter being an antilinear involution. It is assumed, that [D, CP] = 0.
(Violation of CP through K M mixing does not influence the subsequent
arguments which are designed to reveal the symmetry structure of the
representation  of the algebra acting on $ {\cal H} + \overline{{\cal H}}$).

\bigskip

To carry out this program Connes postulates 2 axioms concerning the
involution. Let $ J = CP$ and ${\cal A}$ the algebra. If $ \rho(a)$ is a
representation of $ a (a \in {\cal A})$ one forms the conjugate representation
 $ J \rho(b) J^{-1} (b \in {\cal A})$. Thus the action of the conjugate
representation on $ {\cal H}$ is to first send ${\cal H}$ to $\overline{{\cal
H}}$, then operate with a representation of ${\cal A}$ on antiparticles and
finally send the result of this last operation back on to ${\cal H}$. It is
natural to postulate Axiom I

\begin{eqnarray} \label{18} [ \rho (a), \rho^0 (b) ] \equiv [ \rho(a), J
\rho(b) J^{-1}] = 0, \quad a, b \in {\cal A} \end{eqnarray}

\bigskip

\noindent since $ {\cal H} \oplus \overline{{\cal H}}$ is a direct sum and one
should be allowed to operate on each of its sectors independently. But the
consequence of (18) is far -reaching in that one may then construct ${\cal H}$
which is a direct product representation wherein $\rho(a)$ acts on the first
index of the tensor product and the representation $ \rho^0 (b)$ of the
conjugate algebra acts on the second index (I refrain from the use of the
terms left and right indices to avoid confusion with L, R). Specifically
$\rho^0 (b)$ acts on the first index of $ \overline{\cal H}$. Two equivalent
expressions to write these algebraic actions are

\begin{eqnarray} \label{19} \rho (a) \xi \rho^0 (b) = \rho (a) J \rho (b^*)
J^{-1} \xi \quad ; \xi \in {\cal H}  \end{eqnarray} \bigskip

\noindent In tensor representations of groups, one is accustomed to this
construction -say to make an adjoint representation out of the fundamentals.
What is not so customary is the explicit use  of Axiom 1 to make this
construction. It permits one to use different members of the direct sum which
defines ${\cal A}$ to construct the representations $\rho$ and $\rho^0$. Their
joint action, (19) clearly realizes (18)\underline{ provided one can effect
the 2 operations in arbitrary order}. What Connes realized is that the
involution ${\cal H}\to \overline{\cal H}$ permitted this generalization of
the usual adjoint representation. Thus the bimodular representations (i.e.
having 2 indices) is deeply rooted in the charge conjugation invariance of
physics (i.e. CP). Majorana neutrinos won't do, in this particular
construction. \bigskip

\noindent It is to be noted that this type of representation of an algebra
${\cal A}$ requires that $ {\cal A}$ be a direct sum. If not one would
construct a bimodule which is the adjoint representation of a group and lose
contact with the fundamental notion that the use of algebras restricts the
representations to the fundamentals (Section 2). In using (18)  where  $a$
and $b$ refer to subalgebras which commute, one  retains the linearity
property of the subalgebras. 

\bigskip

 But to exploit this idea, a second axiom is required, which as we shall see
is rooted in the physical requirement of independent gauge fields associated
with the unitaries of the 2 representations $ \rho$ and $ \rho^0$. The axiom
is suggested by the fact that $ [D, \ ]$ is a derivative, a first order
operator obeying a Leibniz rule. It is formulated through Axiom 2

\begin{eqnarray} \label{20} [ \rho (a), [D_M, \rho^0 (b) ]] = 0 \end{eqnarray}

\bigskip

\noindent which from (15), also implies $ [\rho^0 (b), [D_M, \rho (a) ]] = 0$.

\bigskip

Axiom 2 will be seen, in what follows, to deliver the powerful result that
color symmetry is unbroken i.e. quark masses do not depend on their color and
Q C D is a pure vector theory. Such a powerful result therefore calls for a
closer analysis of the physics behind (20). It is gauge symmetry that lurks
in the background. To see this it suffices to present the problem in a more
general  but more familiar setting. Suppose $ \psi$ is bimoledular, a 2 index
entity which is covariant under transformation by 2 groups: $ \psi \to U V
\psi$. (For us $ V = JU^\prime J^{-1}$, but this fact is not cogent to the
present discussion). The gauge principle then must be generalized to cover
this case, and the way one does this is to expand $U$ and $V$ to first order
about unity, expressing thereby each in terms of group generators multiplying
infinitessimal space-time dependent parameters. One then proceeds to ensure
the gauge invariance of the action by inventing two vector potentials, say
$A_U$, and $A_V$ each of which transforms in the usual way for infinitessimal
transformation

\begin{eqnarray} \label{21} \delta A_{\mu, U} &=& - [ A_{\mu,
U},\lambda_{a,U} ]  \in_{a,U}  + \partial_{\mu}\in_{a,U}  \nonumber \\ 
\delta A_{\mu, V} &=& - [ A_{\mu, V}, \lambda_{b,V}]  \in_{b,V} +
\partial_{\mu}\ \in_{b, V} \end{eqnarray}

\bigskip

\noindent Given that $ \delta \psi = \in_{a,U}
 \lambda_{a,U} \psi + \in_{a,V} \lambda_{b,V} \psi$, equation (21) ensures
that $ \delta$(Action)$ = \delta [ \psi^+ (D + A) \psi] = 0$ wherein I have
abbreviated  $A\!\!\!/_U + A\!\!\!/_V$ and $D = i \partial \!\!\!/$. In (21)
the $ \lambda^\prime$s are the group generators and $ \in ^\prime$s are the
infinitessimal parameters of the transformations.

\bigskip

It is then tacitly assumed that this procedure carries over to finite
transformations $ \psi \to U V \psi$ and

\begin{eqnarray} \label{22} A_{\mu, U} \to U A_{\mu, U} U^+ + U \partial_\mu
U^+ \nonumber \\ A_{\mu, V} \to V A_{\mu, V} V^+ + V \partial_\mu V^+
\end{eqnarray}

\bigskip

\noindent But it will be seen that the invariance of the action requires not
only the condition that $ [U, V]$ = 0 but also $ [U, [D, V]] = 0$. Indeed one
has, assuming $ [U, V] = 0$

\begin{eqnarray} \label{23} \delta \Psi^+ D \psi =\psi^+ V^+ [D, V] \psi +
\psi^+ V^+ U^+ [D, U] V \psi \end{eqnarray} \bigskip

\noindent For this variation to be compensated by the gauge potentials
transforming as in (22) then requires this additional rule of commutation. In
short not only must i-spin and color commute, but the variation of one of
them, must be invariant under transformations of the other. \bigskip

Since Connes is taking over the gauge principle in algebraic form for the
gauging of symmetries under displacements $ L \leftrightarrow R$, he then has
little choice but to require (20). What normally is tacitly taken for granted
in $YM$ theory requires this explicitation.

\bigskip

As in Section 4 wherein the matrix $D_M$ led the way towards the dynamics of
SB$\chi$S, we once again are confronted with the novelty of Connes' approach.
The entries in $D_M$ are $c$ numbers, masses, and from the extended gauge
hypothesis, field properties -here those of a scalar field- are deduced. Not
the inverse! The theory so derived is much more constraining. For example, as
we shall see, color is unbroken, quark masses do not depend on their color.
Yet in field theory, nothing in principle prevents such breaking. Scalars can
depend on color as well as flavor and so can their expectation values. But
nature doesn't work that way. As far as internal symmetries are concerned,
she conforms to Connes'axioms. Then one must ask: at bottom, is the Connesian
phenomenology any more outrageous than postulating a certain set of scalar
fields? Indeed given that fermions have masses, it is rather more inductive
than the usual approach -in that it begins ``after the fact".

\bigskip

I now proceed to Connes'constructions. For $ \rho(a)$, the action of ${\cal
A}$ on the first index, he uses the GWS scheme modified so that in
combination with the action on the second index, one recovers the usual
leptonic assignments. It is assumed that for one generation (and once again
it suffices to work with one generation), ${\cal H}$ is $ (\nu_L e_L), e_R,
(u_L, d_L) u_R, d_R$ wherein bracketed fermions are i-spin doublets. That
$u_R$ and $d_R$ are i-spin singlets is an assumption that finds its basis in
the reasoning of the previous section. The action $\rho(a)$ on ${\cal H}$ is
taken to be color blind; one uses $ {\bf C} \oplus \vert\!{\rm H}$ and choses
the phases of operations in both $ \rho$ and $ \rho^0$ to satisfy $Q_R = Q_L$.
\bigskip

\setcounter{equation}{23} \begin{eqnarray} \begin{array}{c|c|c|llllll|c|c|}
     & \nu_L  &   &q& & & &                 &            &\;\; &\nu_L\\
     &  e_L   &   & &  & & &       &                     &\;\;& e_L  \\
     &  e_R   &   & & \lambda^*    & & &    &            &\;\;& e_R  \\ 
\rho(a)&u_L   & = & &           &q & &  &                &\;\;& u_L \\
     & d_L    &   & &           &  & & &                 &\;\;& d_L \\
     & u_R    &   & &           & &      & \lambda&      &\;\;& u_R  \\
     & d_R    &   & &           & &      &      &\lambda^* &\;\;& d_R 
\end{array}  \end{eqnarray} 

\bigskip \noindent In Eq. 24) read in descending order: $q$ is a $ 2 \times
2$ matrix, $ \lambda^*$ is $ 1 \times 1$; $q$ that acts on $(u_L, d_L)$ is
the collection of $C$ matrices each of which is $ 2 \times 2$ and $ \lambda,
\lambda^*$ acting on $u_R, d_R$ resp. are multiplied by unit $C \times C$
matrices. \bigskip

Color is legislated into the theory by taking $ u_L, d_L, u_R, d_R$ to be $C$
dimensional vectors and one uses $  \rho^0 (b)$ to act on the color indices.
Clearly the choice $m (m \in M_c (C))$ for $ \rho^0 (b)$ acting on the quarks
permits one to satisfy $ [ \rho (a), \rho^0 (b) ] = 0$. Another option could
be to take for $ \rho^0 (b)$, the matrix $ \lambda \otimes I$ to operate on
$(u_L, d_l)$ and $M_c$ to operate on $u_R, d_R$. This is however disallowed by
Axiom 2 which takes on the form

\begin{eqnarray} \label{25} [M \rho_R (a) - \rho_L (a) M ] \rho_R^0 (b) = 
\rho^0_L (b) [ M \rho_R (a) - \rho_L (a) M ] \end{eqnarray}

\bigskip

\noindent (for all $a, b!$) wherein we have decomposed $ \rho$ into its chiral
sectors and used the anti-diagonality of $D_M$ in chiral representation.
Applying (25) to the quark sector and using the fact that $ [ M \rho_R (a) -
\rho_L (a) M] \neq 0$ since $ \rho_R \neq \rho_L$ due to flavor splitting it
is seen that $ \rho_R^0 (b) = \rho_L^0 (b)$ and that each commutes with $ [M
\rho_R - \rho_L M]$, Axiom 2 delivers the result that color is a pure vector
theory. Moreover, since $ [M \rho_R (a) - \rho_L (a) M]$ commute with $
\rho^0_R $(or $ \rho^0_L)$ in the quark sector, we have the additional strong
result that this mass breaking term commutes with color.

\bigskip

\noindent As we have seen in Section 4, this matrix is essentially the scalar
field, so that the result can be stated that the scalar field is color
independent. \bigskip

The same reasoning yields that $ \rho^0_R =  \rho^0_L$ on leptons as well.
Furthermore Axiom 1 also implies that $ \rho^0$ acting on $(\nu_L, e_L)$ is
proportional to the unit matrix. Therefore $  \rho^0 (b)$ acting on leptons
is either $ \lambda \otimes I$ or $\lambda^* \otimes I$. The condition $ Q_L =
Q_R$ for $e$ delivers the second option so that, in fine, one gets using the
same convention as previously, the block form

\setcounter{equation}{25} \begin{eqnarray} \rho^0 (b) =
\begin{array}{|ccccccc|} \lambda^*& & & & & &  \\
 & \lambda^*& & & & & \\
  & & \lambda^* & & & & \\
 & & & m& & &  \\
 & & & & m& &  \\
 & & & & & m& \\
 & & & & & & m  \end{array} \end{eqnarray}

\bigskip

One last step is required to complete the construction. The set of unitaries
of $M_c$ is $U_c = U_1 \times S U_c$ (to within irrelevant homotopy). It is
therefore necessary to fix this one last phase. It is to be noted that unlike
$M_2 (C)$ which has $\mid \!\! {\rm H}$ as subalgebra, hence a set of
unitaries $S U_2$, the algebras $M_c$ do not enjoy this property. Thus for $
C \neq 2$ this extra phase is necessarily present. And it is precisely this
phase which permits an adjustement so as to satisfy anomaly freedom: to wit
this phase is $(1/2 C)$. Hypercharge is then obtained from the sum of phases
carried by $ \rho(a)$ and $ \rho^0 (b)$. Reading off (24) and (26), wherein
each entry is in the unitary subset of the corresponding algebraic element,
then gives back Equs. (12) to (16).

\bigskip

Connes makes an important remark concerning this last point. One easily checks
that with the above assignments the total phase of $ \rho(a) \rho^0(b)$
vanishes. [ The count is obtained from 4 factors of $m$ with 4 factors of
$\lambda^*$; the $S U_2$ part of $q$ carries no total phase]. This is what
anomaly freedom has given as result. But Connes remarks that were there a non
trivial total phase, it would be a factor of the algebra which would multiply
the 15$\times$15 unit matrix. Its commutator with $D_M$ would vanish.
Therefore it would not be associated with a scalar gauge potential. Of course
the vanishing of this phase is the unimodular condition. It is then seen that
this total phase has nothing to do with the dynamics coupled to internal
symmetries. This then is another one of these remarkable facts. The algebraic
approach leads to zero coupling of total phase in the internal algebra and
the anomaly structure requires that this phase not be present. There is an
overlap which remains to be understood. The algebraic approach appears to
express ``quantum roots" which are hidden in the formalism. 

\bigskip

In summary, once it is admitted that L-R displacements lead to gauge theory in
the manner of Connes, then the construction of bimodules conforming to the two
axioms on real structures leads to a result in complete conformity with
anomaly freedom wherein color is a pure vector theory (given flavor breaking)
and moreover quark masses independent of color. This latter has nothing to do
with anomalies, but of course is the way nature works.

\vskip1cm

\section{Further Comments} \bigskip

\begin{itemize} \item[a)] Connes has elaborated an elegant topological
argument $^{3)}$ which gives a deeper sense to his bimolecular construction.
It is based on the powerful techniques of $K$ theory developed over the past
few decades and deserves a paper in itself. I shall simply sketch here some
of the ingredients, since at the moment the argument is more one of
consistency then of an independent construction. \bigskip

One forms a matrix $K_{ij}$ where $i$ and $j$ take on values 1, 2, 3
corresponding to the members of $ {\cal A} ( = {\bf C} \; \oplus \mid
\!\!\!{\rm H} \oplus M_c)$ (i.e. the subalgebras span a vector space).
$K_{ij}$ receives a contribution, which is calculated according to a rule of
projection. It is non vanishing when there are one or more members of ${\cal
H}$ whose first index transforms according to ${\cal A}_i$ in $ \rho (a) $[
where $ {\cal A}_1 = {\bf C}_1; {\cal A}_2 = \mid \!\!{\rm H} ]$ and whose
second index transforms according in ${\cal A}_j$ in $ \rho^0 (b)$ [ where
${\cal A}_1 = {\bf C}, {\cal A}_3 = M_c]$ and vice versa. Thus $ K_{ij}$ is a
sort of measure of whether or not there are common attributes in the two
indices that make up the bimodule. It is also important that the sign of the
contribution change with chirality. The theorem is that if $ \vert K \vert
\neq 0$, the manifold on which is built the algebra is a topologically
acceptable space. If it is singular, the whole bimodular algebraic
construction makes no sense. For S M with Connes'assignments it works. If one
tries to add in $\nu_R$ where the latter has zero charge and zero i-spin, one
finds $ \vert K \vert = 0$. Upon performing the calculation one sees that to
have $ \vert K \vert \neq 0$ requires L, R dissymmetry in ${\cal H}$ of the
GSW type. Otherwise the columms which make up $K_{ij}$ become linearly
dependent.

\bigskip

The above result is satisfying since such a $ \nu_R$ would decouple from both
the  YM fields and the scalar field of SM.  Unfortunately there has been no
systematic study on what would be other acceptable models based on this
critereon. Nevertheless the above calculation does suggest that parity
violation is essential. And this is in keeping with remarks I made in Section
3. \bigskip

It is rather important that this question be studied more extensively. For if
my conjecture is borne out, parity violation would be elevated to a dynamical
principle, based on topological arguments.

\bigskip

\item[b)]  Similar schemes have been developed by other authors $^{17),
18)}$. I have not reviewed them here, not because of their lack of interest,
but rather because only Connes'scheme has been developed in sufficient
detail, and appears sufficiently constrained, to make contact with
phenomenology. In particular there is now a body of quantitative work which I
briefly mention in point c) below.

\bigskip

\item[c)] Chamsedinne and Connes 6) have calculated an effective action for
bosonic fields as follows. One first constructs the full Dirac operator
$D_{Tot}$ in the presence of gravity, YM and Higgs fields. It is possible to
introduce a set of coupling constants which is consistent with all the
algebraic constraints so that $D$ is of the form used in SM phenomenology.
These coupling constants are obtained by introducing different weights to
various independent sectors of the total trace that makes up scalar products
$^{19)}$. The weight factors are diagonal matrices which commute with the
matrices that represent ${\cal A}$.

\bigskip

The calculation then consists in the evaluation of the number of states whose
eigenvalue of $ D^2 ( = \lambda^2)$ is such that $ \lambda^2 < \Lambda^2$ when
$ \Lambda$ is a cut-off parameter and one works in the euclidean domain. The
evaluation is made in descending powers of $ \Lambda$ and the first three
terms are retained these being considered ``relevant". For flat space only
the $ \Lambda$ independent term contributes, whereas in curred space one has 
a $ \Lambda^4$ cosmological constant $ + \Lambda^2 \times$ (Einstein-Hilbert
action). Note that the calculation is entropistic in character and not the
same thing that one gets on integrating over fermi fields (as has sometimes
been done in an effort to generate gravity dynamically). Indeed to the above
bosonic action one adds the fermionic action.

\bigskip

It is interesting to remark that the Bekenstein-Hawking (BH) entropy which
arises in the presence of event horizons is related to the Connes-Chamedinne
(CC) effective action, calculated as it is, in the eucidean. Indeed when the
BH entropy is calculated as a functional integral over fields in a periodic
domain, the integrand is the exponential of the total action. And this
according to CC is obtained from a count of states. Thus, in this vision the
BH entropy is the average over one cycle of the CC count of states. Of course
for this to be truly useful one will have to supplement CC by a viable theory
of gravity.

\bigskip

What is interesting in the present context is that in recent calculations
$^{20)}$ carried out in flat space, the renormalization group, which shows how
all couplings run with $\Lambda$, yields the acceptable result, that given
their present experimental values one reaches symmetry (electroweak and
strong) at $ \Lambda$ about 10$^{15}$ GeV. Moreover if one uses the value of
the top mass $( \sim$ 175 GeV), one predicts a Higgs mass of $ \sim$ 200 GeV.
\bigskip

Previous calculations along the same lines $^{19)}$, but based on the
construction sketched in Section 3 gives similar results, but according to the
authors, accord less well with certain experimental facts.

\bigskip

It also has been pointed out to me by T. Schucker that the flexibility induced
by the choice of weights in the trace when color is present destroys the no-go
theorem of Section 4. It remains to be seen how this statement should be
interpreted when the renormalization group is applied so as to reach the
scale of symmetry restoration since then the weights are once more all the
same. Thus one must reserve judgement as to whether the scheme does or does
not require more than one generation.

\bigskip

As previously mentioned, there is at present no explanation of why there are
three generations. Nor is these any indication in the scheme to explain the
widely disparate mass scales from generation to generation, nor the KM mixing
and CP violation. One must also be prepared for the eventuality that the Higgs
scalar is not a local field. The whole phenomenology may break down at some
intermediate large mass scale. As I have emphasized Connes'construction
provides a rational framework in which to set SM phenomenology. Many facts
fall into place which seem to have no rhyme or reason, and in this sense I
consider it significant.

\bigskip

\item[d)] With regard to gravity, since the symmetry value of $ \Lambda
(\equiv \Lambda_s)$ is less than $m_{planck}$ by some orders of magnitude, it
would seem that the model used to construct $D$, (i.e. the usual continuum
approach of Riemann geometry) must be modified at larger mass scales than
$\Lambda_s$. This is quantum gravity - the formulation of which is the
primary goal of present- day physics. At present we are still far from this
goal. Nevertheless it is suggested that the reader look into refs $^{5), 6)}$
which contain considerations on automorphisms of the algebra. Those which are
norm conserving are concerned with internal symmetries and the others with
gravity. This novel approach may help point the way. Also, Connes has pointed
the possible relevance of  quantum groups $^{5)}$. (The first extension of
groups to quantum groups contains the algebra $ {\cal C} \oplus M_2 ({\cal
C}) \oplus M_3 ({\bf C})$ - Of all things!).

\item[e)] It has been shown by Lizzi$^{20)}$ and collaborators that unless one
adds new fermions, it is not possible to construct GUTS schemes which conform
to Connes'axioms. It appears that SM is the unique solution.

\end{itemize} \vskip1cm

\section{Acknowledgements} \bigskip

This is to express my profound gratitude to Alain Connes for his
explanations. I hope that these Notes reflect in some measure his scientific
acumen and imagination. \bigskip

This paper also reflects conversations with many colleagues whom I here thank
with pleasure: F. Bastionelli, R. Coquereaux, J.-M. Fr\`ere, B. Jochum, T.
Krajewski, T. Schucker, D. Testard and above all D. Kastler who not only has
offered me gracious hospitality but who has taken upon himself the difficult
task of teaching me operator algebras. Section 5 on anomalies was instigated
by a suggestion of J.-M. Fr\`ere and implemented by important remarks of F.
Bastionelli who showed me how the condition $Q_L = Q_R$ led to a clear
understanding of internal consistency. Bastionelli also alerted me to the
consistency with the gravitational anomaly.

\vskip1cm \centerline{\bf REFERENCES}

\begin{itemize} \item[1)] A Connes in The interface of Mathematics and
Physics  D.G. Quillen, B.B. Segal and S.T. Tsou, eds Clarendon Press Oxford
(1990). \item[2)] A. Connes and J. Lett. Nucl. Phys. B (Proc Suppl.)
\underline{18} (1990), 29. \item[3)] A. Connes \underline{Non Commutative
Geometry}, Academic Press, London (1994). \item[4)] A. Connes, J. Math. Phys.
\underline{36} (1995) 6194. \item[5)] A. Connes, hep-th/9603053. \item[6)]
A.H. Chamsedinne and A. Connes, hep-th/9606001. \item[7)] F. Englert, R.
Brout and M.F. Thiry, Nuov Cim \underline{43}, 244 (1966). \item[8)] For a
good review see M. Veltman in the forthcoming proceedings of the History of
the Standard Model Stanford University, 1992. \item[9)] F. Englert, R. Brout,
Phys. Rev. Lett. \underline{13}, 32 (1964). \item[10)] P. Higgs, Phys. Rev.
Lett. \underline{13}, 508 (1964). \item[11)] T. Sch\"ucker, J-M Zylinkski, J.
Geom. Phys. \underline{16} (1995), 207. \item[12)] A good review is C.P.
Martin, J.M. Garcia-Bondia, J.C. Varilly, hep-th/9605001. \item[13)] B.
Iochum, T. Sch\"ucker, Lett. Math. Phys. \underline{32} (1994), 153.
\item[14)] C. Misner, K. Thorne and J. Wheeler, \underline{Gravitation}, W.H.
Freeman, New York (1970). \item[15)] C. Nash, S. Sen \underline{Topology and
Geometry for Physicists}, Academic Press, London (1983). \item[16)] See for
example W. Greiner and B. M\"uller, \underline{Gauge Theory of Weak
Interactions}, Springer, Berlin 1993. \item[17)] M. Dubois-Violette, R.
Kerner and J. Madore, J. Math. Phys. 31 (1990) 316. \item[18)] R. Coquereaux,
G. Esposito-Farese and G. Valiant, Nucl. Phys. \underline{B 353} (1991), 689.
\item[19)] B. Iochum, D. Kastler and T. Sch\"ucker, hep-th/9507150. 
\item[20)] Reported by T. Sch\"ucker at the Marseille, Workshop on NCG, March
1997. \item[21)] Reported by F. Lizzi at the Marseille Workshop on NCG March
1997. \end{itemize}

\end{document}